\documentclass[submission,copyright,creativecommons]{eptcs}
\usepackage[T1]{fontenc}

\usepackage{breakurl}  
\usepackage{underscore}
\usepackage{graphicx}
\usepackage{listings}
\usepackage{makecell}
\usepackage{resolve}
\usepackage{mdframed}
\usepackage{mathpartir}
\usepackage{amsmath}
\usepackage{amsthm}
\usepackage{tabularx}
\usepackage{booktabs}
\usepackage{multirow}
\usepackage{wrapfig}

\newtheorem{definition}{Definition}[section]

\theoremstyle{definition}
\newtheorem{example}{Example}[section]

\newcommand\ctxsep{
	\ \text{\textbackslash} \
}

\newcommand{\infrule}[3]{\ensuremath{\mbox{ $\mathsf{#1}$ } \ \frac{\textstyle #2}{\textstyle #3} } }

\title{F-IDEs with Features and VCs Designed to Assist Human Reasoning When Verification Fails}

\author{Yu-Shan Sun
\institute{Clemson University\\
Clemson, SC, USA}
\email{yushans@clemson.edu}
\and
Daniel Welch
\institute{Pennsylvania State University\\
University Park, PA, USA}
\email{dtw5246@psu.edu}
\and
Murali Sitaraman
\institute{Clemson University\\
Clemson, SC, USA}
\email{msitara@clemson.edu}
}

\providecommand{\event}{AppFM 2021}
\def\titlerunning{F-IDEs Designed to Assist Human Reasoning When Verification Fails}
\def\authorrunning{Y. Sun et al.}

\begin{document}
\maketitle

\begin{abstract}
This paper summarizes our efforts to aid human reasoning when verification fails through the use of two distinct Formalization Integrated Development Environments (F-IDEs) that we have developed. Both environments are modular and facilitate reasoning about the full behavior of object-based code. The first environment, referred to as the web-IDE, has been used for several years to teach aspects of formal specification and verification, including why and where verification conditions (VCs) arise and how to use them when verification fails. The second F-IDE, RESOLVE Studio, remains experimental, but is a more fully-fledged environment backed by a sequent-based VC generator that produces VCs with fewer extraneous givens. While the environments and VC generation techniques are necessarily language specific, the principles of alternative VC generation methods, F-IDE features, and observations about their impact on novices and experienced users are more generally applicable.
\end{abstract}

\section{Introduction}
\label{sec:intro}

As the importance of tools and environments with features for supporting software verification is becoming better understood, a variety of systems have been developed to fill this need. Indeed, the usability of \textit{auto-active}~\cite{leino-autoactive:2010} specification and verification languages such as Why3~\cite{fillatre:2013}, Dafny~\cite{leino:2013}, RESOLVE~\cite{sitaraman:2011}, and AutoProof~\cite{tschannen:2015}---in which users indirectly interact with automated provers through formal contracts such as loop invariants---hinge almost entirely on the feedback provided to users through their respective front-end environments. 

Auto-active style feedback typically takes the form of a collection of necessary and sufficient verification conditions (VCs) for proving correctness of code w.r.t. some formal specification. Different tools report (failed) VC details differently. For example, Dafny tends to report assertion failures through Z3-generated counter examples~\cite{demoura:2008}, whereas AutoProof provides a higher-level English explanation for each VC. While each tool has its distinct characteristics, they share the common need of providing effective, non-ambiguous support to users (preferably of all experience levels) when verification fails.

The F-IDE contributions of this paper fall under the categorizations of \emph{usefulness} and \emph{ease of use}. We consider how VC usefulness is achieved and how it can be effective. We detail how ease of understanding for a VC is achieved, focusing, for example, on how reasoning complexity is simplified for beginners by minimizing givens.

When necessary and sufficient VCs are generated and proved correct, their details are understandably of little consequence to software engineers. So why focus on VC generation at all? We consider two reasons. The first is practical: when verification fails, VCs represent a first, crucial foothold for novices and experts alike when attempting to identify which fixes are needed at the source code or specification level. The second reason is pedagogical: even when verification succeeds---students should be able to examine the details of a VC, understand which line of code (or construct) generated it, and why it was provable (e.g., making connections to concepts learned in discrete math courses). Consequently, easing the human reasoning process (through simpler, smaller VCs) and supporting student inquiry into what happens when a program is verified are chief motivations for the technical VC generation enhancements and tool features discussed in this paper.

We highlight aspects of two F-IDEs that we have built as the primary means of viewing VCs, proving, and debugging code that has failed to verify. The language targeted by both environments is RESOLVE~\cite{sitaraman:2011}---an integrated specification and programming language that is imperative and object-based. The language---with variants adapted to popular languages such as C++ and Java~\cite{heym:2017}---has been used to teach modeling and reasoning principles (such as Design by Contract) from beginning undergraduate CS education to graduate level courses. Over 25,000 students across multiple institutions have benefited from from these efforts over two decades~\cite{cookicse:2012,heym:2017,kabbani:2015}. 

\begin{figure}[!t]
\centering
\includegraphics[scale=.95]{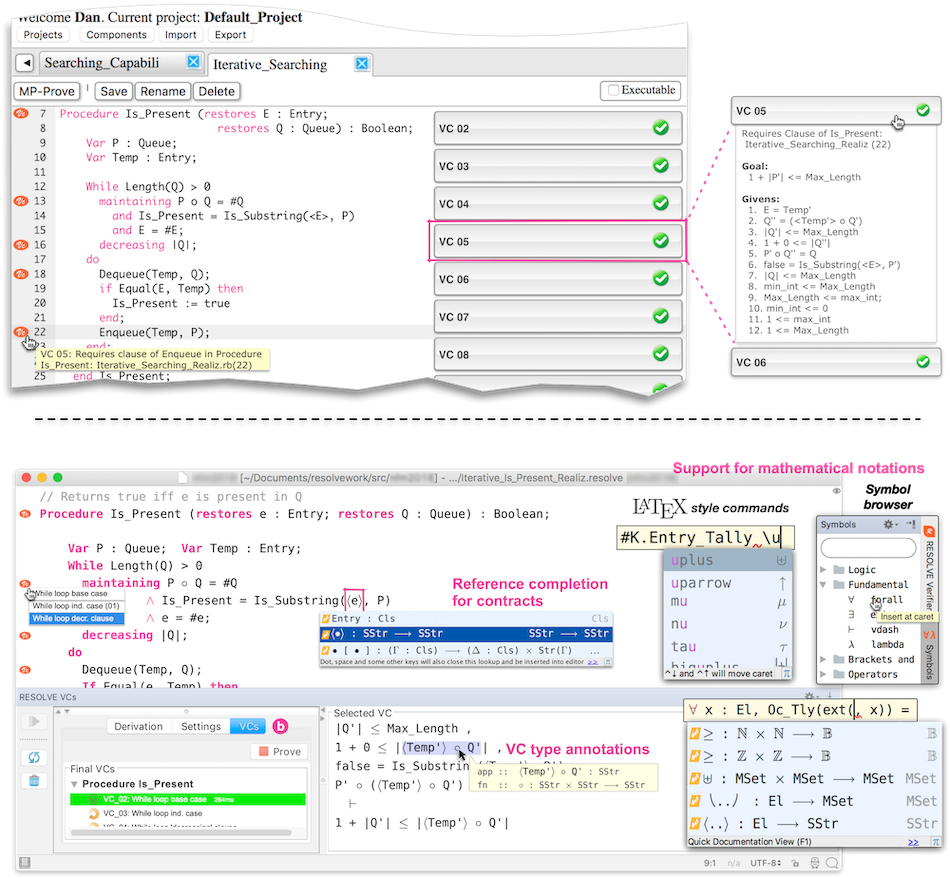}
\caption{RESOLVE's web-based F-IDE (top), and RESOLVE Studio (bottom).}
\label{fig:fides}
\end{figure}

The older of the two environments is the web-IDE (Fig.~\ref{fig:fides}, top), which has been in use for over a decade as the primary development environment for RESOLVE.\footnote{\url{https://resolve.cs.clemson.edu/teaching}} The second, more recent F-IDE is a desktop based environment named RESOLVE Studio (Fig.~\ref{fig:fides}, bottom). Built on top of the JetBrains IDE platform,\footnote{\url{https://www.jetbrains.com/}} RESOLVE Studio combines the usual modern IDE amenities with a version of the RESOLVE compiler that includes the revised VC generation scheme detailed in this paper. Additionally, in an effort to guide the design and application of proof rules, RESOLVE Studio allows users to interactively derive VCs from Hoare style triples. While this environment remains under development, it has seen some usage (albeit limited) in a graduate programming languages course to construct specifications using small, student-defined mathematical theories. The contributions of this paper are twofold:
\begin{itemize}
\item A general demonstration of how specification and auto-active verification of software can be supported in the context of two F-IDEs. 

\item A revised set of proof rules for generating sequent-based VCs in a \textit{parsimonious} manner that omits irrelevant givens with the intent of improving human reasoning and VC comprehension.
\end{itemize}

The paper is organized as follows. Section~\ref{sec:background} provides background on the RESOLVE language. Section~\ref{sec:motivation} illustrates the need for simpler VCs by examining a failed VC in an older version of our web-IDE. Section~\ref{sec:vcgrules} provides a high level overview of RESOLVE's VC generation process, while section~\ref{sec:proofrules} details the proof rules that enable generation of parsimonious VCs. Section~\ref{sec:resolvestudio} shows VC feedback and other features in RESOLVE Studio. Sections~\ref{sec:expeval} -~\ref{sec:conclusion} contain respectively: an experimental evaluation of the revised proof rules, related work, and conclusions with directions for future work.

\section{RESOLVE Background}
\label{sec:background}

RESOLVE~\cite{sitaraman:2011} is an imperative, object-based programming and specification framework designed to support modular verification of sequential code. Every RESOLVE component has a formal interface specification called a \lstresolve|Concept|. Below is a snippet of one such concept for bounded queues.

\begin{lstlisting}[language=resolve]
Concept Queue_Template (type Entry; evaluates Max_Length : Integer);
  requires Max_Length > 0;
  uses String_Theory;
  
  Type family Queue is modeled by Str(Entry);
    exemplar Q;
    constraints |Q| \leq Max_Length;
    initialization ensures Q = $\Lambda$; // note: Q = \Lambda implies |Q| = 0
\end{lstlisting}

The concept shown is parameterized by a generic type \lstresolve|Entry| and an integer \lstresolve|Max_Length| that places an upper bound on the number of entries that can be stored. This bound must be positive (as per the module level \lstresolve|requires| clause) and may be an arbitrary expression of type \lstresolve|Integer| (as per the \lstresolve|evaluates| parameter mode). The imported module, \lstresolve|String_Theory|, gives the concept access to a mathematical theory of strings---including operators for: string length $|\bullet |$, concatenation $\circ$, the empty string $\Lambda$, and a singleton string formation function $\langle \bullet \rangle$.

\subsection{Abstract Specification}

Specifications in RESOLVE are \textit{model-based}: that is, each programmatic type is conceptualized through the \lstresolve|Type family| construct, which, in this case, declares that the (program) type \lstresolve|Queue| is mathematically modeled as a string of generic entries, i.e.: \lstresolve|Str(Entry)|. The \lstresolve|exemplar| queue, \lstresolve|Q|, that follows gives specifiers a formal name to an example queue within the declaration. It is used immediately thereafter in the \lstresolve|constraints| to assert that (1) not \textit{all} strings are models of valid queues, but only those of length \lstresolve|Max_Length| or less, and (2) that queues are empty upon \lstresolve|initialization|.

Operations are declared after the model. Below is one such operation for \lstresolve|Enqueue|. 
\begin{lstlisting}[language=resolve]
  Operation Enqueue (alters e : Entry; updates Q : Queue);
    requires |Q| < Max_Length;
    ensures Q = #Q $\circ$ $\langle$#e$\rangle$;
\end{lstlisting}

Formal contracts for \lstresolve|Enqueue| are communicated through \lstresolve|requires| and \lstresolve|ensures| clauses (i.e.: pre- and post-conditions). The contract also encompasses \textit{specification parameter modes}. A mode of \lstresolve|updates| means that while the value of the incoming queue (\texttt{\#Q}) may be meaningful, its outgoing value (\texttt{Q}) is to be updated in the manner specified in the \lstresolve|ensures| clause. The \lstresolve|ensures| clause can be read as follows: \textit{``the outgoing value of Q is equal to the \#-denoted incoming queue concatenated with the singleton string containing the incoming value of entry e.''} The \lstresolve|alters| mode means that the incoming \texttt{\#e} may contain a meaningful value, but its outgoing value \texttt{e} is left unspecified. This gives implementers of the \texttt{Enqueue} operation the most flexibility as to which value is stored in the parameter \texttt{e} after a call to \texttt{Enqueue} completes.

To facilitate modular reasoning, the language enforces a strict separation between the abstract state expressed in interface specifications and executable, implementation level code. Thus, implementations must provide the correspondence information (abstraction functions or relations) necessary to connect the concrete state (in the implementation) to the abstract state (in the interface). Readers interested in a more complete description of RESOLVE should consult\sloppy{~\cite{smithphddiss:2013,hartonphddiss:2011,sitaraman:2011,sunphddiss:2018}} or one of the case studies~\cite{mbwambomsthesis:2017,tagore:2012,welch:2017} carried out using the language.

\section{The Need for Simpler VCs: Examining a Failed VC}
\label{sec:motivation}

While even modest enhancements made to our presentation of VCs (e.g., annotating the source of VCs next to the line(s) on which they arise) have helped, the complexity of debugging VCs for beginners is considerably increased when they include a large, obfuscating number of irrelevant givens. 
\begin{figure}[!htb]
\centering
\includegraphics[scale=1.05]{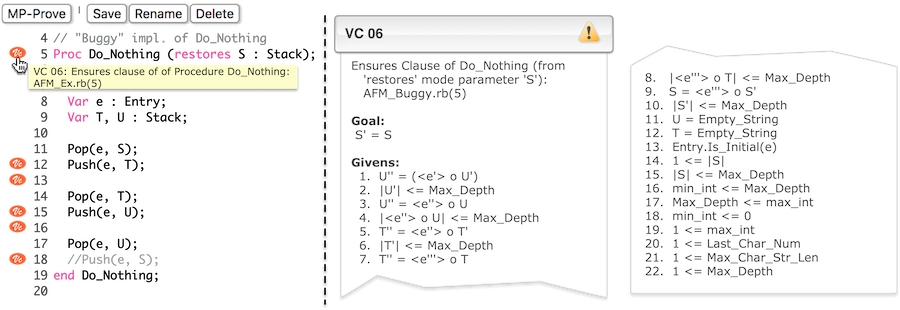}
\caption{A VC that fails to prove that stack \texttt{S} is restored; note: the version of RESOLVE pictured does not accept non-ASCII notations---so \lstresolve|Empty_String| is used instead of $\Lambda$, \texttt{<..>} instead of $\langle..\rangle$, etc.} 
\label{fig:failedvc}
\end{figure}

For illustration purposes, Fig.~\ref{fig:failedvc} shows a buggy implementation of a \lstresolve|Do_Nothing| operation on stacks\footnote{Stacks and queues are modeled the same---though their operations (and specs.) differ} which \lstresolve|requires| \texttt{|S| >= 1}, but fails to verify since the last call to \lstresolve|Push| is commented out. In this case, the failed VC was attempting to prove that \texttt{S} had been restored to its original state---as per the \lstresolve|restores| parameter mode on \texttt{S} which automatically adds \texttt{S = \#S} as a conjunct to \lstresolve|Do_Nothing|'s \lstresolve|ensures| clause. 
After pressing the \textbf{\textsf{MP-Prove}} button, VCs are generated for pre-conditions of invoked operations, in addition to the post-condition of the operation being proved. Givens added to each VC arise from module level \lstresolve|requires| clauses and external \lstresolve|constraints| from imported modules.

The process is completed when all VCs are verified or the system times out. Since many initial student attempts contain errors and fail to verify, to provide quicker feedback, the timeout has been set to be minimal on the web-IDE at the risk of not being able to discharge some otherwise provable VCs. Further, once at least three VCs fail to prove, the other VCs are not attempted. Hovering over the orange VC badges on specific lines gives information about the source of the VC(s) that arise from that line.

\section{VC Generation Process}
\label{sec:vcgrules}

Generation of VCs that are both necessary and sufficient in order to prove that an implementation is correct w.r.t. its specification is a syntax-directed process. A more detailed description of RESOLVE's VC-derivation scheme and its proof rules can be found in~\cite{hartonphddiss:2011,sunphddiss:2018}.

Prior to the application of any statement level proof rules is a pre-processing step in which user code is logically grouped into \textit{assertive-code} blocks wherein all mathematical assertions are made explicit. The traditional Hoare triple of the form $\{P\} \ c \ \{Q\}$ looks like the following in our presentation, as assertive code may include any number of preceding statements or assertions such as \texttt{\textbf{Assume}} $P$:
\begin{align*}
\mathcal{C} \ctxsep \ \mathit{c}; \quad \texttt{\textbf{Confirm}} \ Q;
\end{align*}

Here, $\mathcal{C}$ is a context containing a collection of typed symbols, $\mathit{c}$ is a sequence of zero or more program statements, and $Q$ is an assertion that must be confirmed to hold at the end. 

Throughout the remainder of the paper, we employ a \textsf{san-serif style font} when typesetting the names of proof rules, as well as any meta-operators we define along the way.

\subsection{Abstract Syntax}
\label{sec:abstractsyntax}

RESOLVE's specification language is built on a many-sorted first order logic organized into several constituent parts. First, we assume an initial set of sort symbols $\mathsf{S} = \{ \mathtt{SSet}, .. \} \cup \{ \mathbb{B} \}$. Here, \texttt{SSet} is the proper class of sets and the notation $T : \mathtt{SSet}$ introduces a new user defined sort $T$ into context.
\begin{itemize}
\item A set $\mathcal{T}$ of nullary constants $c_1, .., c_n : T \in \mathcal{T}$ where $c_1, c_n$ could also have any other sorts in $\mathsf{S}$. 
\item A set $\mathcal{X}$ of variables $x, y, z : T \in \mathcal{X}$ representing arbitrary elements which range over a domain $T$.
\item A set $\mathcal{F}$ of typed function symbols $f, g, .., h : T_1 \times .. \times T_n \rightarrow T$ which range over $\mathcal{F}$. Here  $T_1 \times .. \times T_n$ represents the domain, while the (non-subscripted) $T$ represents the co-domain. 
\item Lastly, a set $\mathcal{P}$ of predicate symbols $p, q, .., r : T_1 \times .. \times T_n \rightarrow \mathbb{B}$; which also includes a reserved binary predicate for equality ($=$) as well as nullary predicate symbols for \textit{true} and \textit{false}. We denote the arity of a given function or predicate $o$ as $\mathsf{ar}(o)$.
\end{itemize}
These sets constitute the language's vocabulary $\mathcal{V} = \mathcal{T} \cup \mathcal{X} \cup \mathcal{F} \cup \mathcal{P}$; where $\mathcal{V}$ can be enriched via the definition of new constants, functions, and predicates in RESOLVE's object theories. As a result, these sets are not necessarily disjoint in practice. So determining whether a nullary ``symbol'' denotes a variable (e.g, bound under a quantifier) or the name of a module level definitional constant is dependent on the current context, $\mathcal{C}$. With these categories fixed, we now define formulas (which denote truth values) and terms (which serve as the fundamental building blocks of formulas). 
\begin{definition}
The set of formulas and terms of our specification language over vocabulary $\mathcal{V}$ is given by the following abstract syntax.
\begin{gather*}
\mathbf{Form}_{\mathcal{V}} \ni \phi, \psi ::= P(t_1, .., t_{\mathsf{ar}(P)}) \ |\  \mathit{true} \ |\ \mathit{false} \ | \ \neg \phi \ |\  \phi \circ \psi | \ \mathcal{Q}\bar{x}_n, \phi \ | \ t\\[.4em]
\mathbf{Term}_{\mathcal{V}} \ni t, y, \tau ::= \ t_0 (t_1, .., t_{\mathsf{ar}(t_0)}) \ | \ t \rightarrow y \ | \ t_0 \times .. \times t_n \ | \ \lambda \bar{x}_n, t \ | \ \cdots \ | \ (\psi) \ | \ \#? \ s
\end{gather*}
where $\circ \in \{\wedge, \ \vee, \ \Rightarrow, \ \Leftrightarrow\}$ and $\mathcal{Q} \in \{\forall, \ \exists, \ !\exists \}$.
\end{definition}
Formulas consist of the usual binary connectives and quantifiers while terms permit (respectively) function application,\footnote{Outfix and infix style applications are also accepted, though we omit these for brevity} function (arrow type) constructors, Cartesian products ($\times$), lambda abstraction, parenthesized formulae, and terminal symbols $s$. We use the notation $\bar{x}_n$ as shorthand for a list of variable binders: $x_1, .., x_n : \tau$ of term/type $\tau$.\footnote{Note that this effectively makes the  specification language higher order, as one can write, e.g., $\forall f : T \rightarrow T, f(x)$; we forego  discussion of this and its ramifications on proof automation---as this paper is primarily concerned with tools and VC-generation}

Next, we establish our syntax for assertive code fragments.
\begin{definition}
\label{def:asrtcode}
A fragment of assertive code consists of zero or more program statements interleaved with assertive statements of the form:
\begin{align*}
&\mathbf{Stmt}_\mathcal{V} \ni s ::= \textnormal{\texttt{\textbf{Assume}}} \ \phi; \ | \ \textnormal{\texttt{\textbf{Confirm}}} \ \phi; \ | \ \textnormal{\texttt{\textbf{Stipulate}}} \ \phi; \ | \ v := t; \ | \ \cdots \ | \ \mathit{id} (y_1, .., y_n);\\
&\qquad \quad \mathbf{AsrtCode} \ni a ::= \ s^* \ \ \textnormal{\texttt{\textbf{Confirm}}} \ \bigwedge seq^+ ; \qquad \mathbf{Seqnt} \ni seq ::= \Gamma \vdash \Delta
\end{align*}
where $\Gamma$ and $\Delta$ are sets of well-formed-formulas (wffs).
\end{definition}
The statement production rule admits verification language specific statements (including \textbf{\texttt{Assume}} and \textbf{\texttt{Confirm}} clauses---which we discuss in Sect.~\ref{sec:proofrules}) as well as strictly programmatic ones such as function assignment ($v := t$) and procedure calls ($\mathit{id}(y_1, ..,y_n)$). We omit any remaining programmatic statements for brevity.

Lastly, since our parsimonious rules rely on the ability to compute sets of free variables from specifications, we employ a contexted ``specification free variable'' function $\mathsf{SFV} : \mathcal{C} \times \mathbf{Term}_{\mathcal{V}} \rightarrow \wp(\mathbf{Term}_{\mathcal{X}})$. The following example demonstrates the function's behavior.

\begin{example}[\textsf{SFV}]
After processing theory modules for mathematical Booleans, Integers, and Strings, suppose $\mathcal{C}$ contains the following constants and predicates:
\begin{align*}
\mathcal{C} = \{ \Lambda : \mathtt{SStr}, \ \wedge :  \mathbb{B} \times \mathbb{B} \rightarrow \mathbb{B}, \  |\bullet | : \mathtt{SStr} \rightarrow \mathbb{N}, \ \mathtt{0} : \mathbb{Z}, \ \cdots \ \}.
\end{align*}
The meaning of the types assigned to the constants should be reasonably familiar with the exception of \texttt{SStr}---which is a user-defined sort representing the proper class of all heterogeneously typed (math) strings. Next, we enrich $\mathcal{C}$ with the following operation signature, where the programmatic type \lstresolve|Static_Array| is mathematically modeled by \lstresolve|\Int \rightarrow Entry|.
\begin{lstlisting}[language=resolve]
$\mathcal{C}' = \mathcal{C} \cup \{ \ $Oper Op (e : Entry; Q : Queue; A : Static_Array);
           ensures Q \neq \Lambda \and e = #A(0) \and 
               A = \lambdaj : \Int, if |Q| = j then #e else #A(j); $\}$$\ctxsep$
\end{lstlisting}
Denoting the \lstresolve|ensures| clause as $\psi$, then $\mathcal{C}' \ctxsep \mathsf{SFV}(\psi)$ yields: $\{ \texttt{Q}, \ \texttt{\#A}, \ \texttt{A}, \ \texttt{\#e}, \  \ \texttt{e} \}$. Note that constants like  $\Lambda$, $| \bullet |$, etc. and bound variables \texttt{j} were excluded, while specificational (program) variables (e.g. \texttt{e}, \texttt{\#e}) were included.
\end{example}

\subsection{Sequent Calculus Review} 
\label{sec:sequentcalculus}

Formally, the final \texttt{\textbf{Confirm}} assertion (which always terminates a fragment of assertive code) is represented as a conjunction of Gentzen-style sequents $\textbf{G}_1, \cdots, \textbf{G}_n$, each of which constitute the final VCs produced by our goal-directed program proof process. 

Each \textbf{G}entzen sequent has the form $\mathbf{G} \equiv \varphi_1, \cdots, \varphi_m \vdash \psi_1, \cdots, \psi_n$ where $m$ and $n$ are non-negative and $\varphi_1,..,\varphi_m$, $\psi_1,..,\psi_n$ denote sets of wffs within a given sequent's \textit{antecedent} ($\Gamma$) and \textit{succedent} ($\Delta$), respectively. The wffs in these sets are added from different specification contexts based on our proof rules. Each sequent \textbf{G} semantically adheres to the usual interpretation: $\bigwedge_{i=1}^m \varphi_i \vdash \bigvee_{j = 1}^n \psi_j$.
\begin{figure}[!htb]
\vspace{-.05cm}
\centering
\begin{mathpar}
\infrule{NotLeft}{\Gamma \vdash \phi, \ \Delta}{\Gamma, \ \neg\phi \vdash \Delta}
\and
\infrule{AndLeft}{\Gamma, \ \phi, \psi \vdash \Delta}{\Gamma, \ \phi \wedge \psi \vdash \Delta}
\and
\infrule{OrLeft}{\Gamma, \ \phi \vdash \Delta \quad \Gamma, \ \psi \vdash \Delta}{\Gamma, \ \phi \vee \psi \vdash \Delta}\\

\infrule{NotRight}{\Gamma, \ \phi \vdash \Delta}{\Gamma \vdash \neg\phi, \ \Delta}
\and
\infrule{AndRight}{\Gamma \vdash \phi, \ \Delta \quad \Gamma \vdash \psi, \ \Delta}{\Gamma \vdash \phi \wedge \psi, \ \Delta}
\and 
\infrule{OrRight}{\Gamma \vdash \phi, \psi, \ \Delta}{\Gamma \vdash \phi \vee \psi, \ \Delta} 
\end{mathpar}
\caption{Standard sequent reduction rules for $\neg$, $\wedge$, and $\vee$.}
\label{fig:reduction}
\end{figure}

To simplify the VCs, we apply the standard sequent reduction rules shown in Fig.~\ref{fig:reduction} to the antecedents and succedents of the sequents that make up the final \texttt{\textbf{Confirm}} until they contain only atomic formulas.

We also fix convenience functions $\mathsf{SFV}_{\mathbf{set}}(\mathit{S})$ which extracts the sfv set from the wffs in a given set $S$, and $\mathsf{SFV}_{\vdash}(\mathit{seq})$ which extracts the sfv set from each side of a given sequent, \textit{seq}.

\subsection{Goal-Directed Proof Rule Application}
\label{sec:proofruleapp}

\begin{wrapfigure}{r}{0.575\textwidth}
\vspace{-0.3in} 
\centering
\includegraphics[scale=.55]{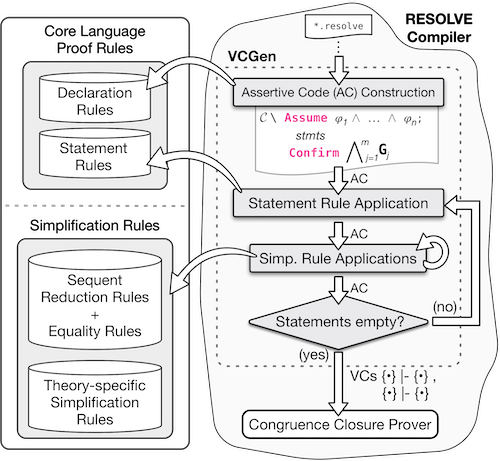}
\caption{RESOLVE's VC generation scheme.}
\label{fig:flow}
\end{wrapfigure}

Once assertive code has been constructed, the approach we use to generate VCs is \textit{goal-directed} and is illustrated at a high level in Fig.~\ref{fig:flow}. 

Starting with the penultimate statement (prior to the final \texttt{\textbf{Confirm}} conjunction), each statement is eliminated one at a time via the application of its corresponding proof rule. After each statement rule is applied, sequent reduction rules and others (such as rules for handling equality or theory-specific rewrite rules) are then applied to further simplify the formulae in each sequent.

After all statements have been eliminated via application of their corresponding rules, the conjuncted sequents in the final \texttt{\textbf{Confirm}} are broken apart and sent off to RESOLVE's in-house congruence closure prover for verification~\cite{smithphddiss:2013,kabbani:2015}. The particulars of the prover are not directly relevant to this paper. Prover backends with which we have experimented include specialized decision procedures~\cite{adcockphddiss:2010} as well as SMT solvers~\cite{tagore:2012} based on Z3~\cite{demoura:2008}.

\section{Proof Rules for Forming Parsimonious VCs}
\label{sec:proofrules}

In this section, proof rules for generating \textit{parsimonious} VCs are formalized and illustrated on a fragment of example assertive code. As with the sequent reduction rules, in the proof rules that follow, it is necessary and sufficient to prove what is above the line in order to prove what follows below the line.

\subsection*{Confirm Rule}
\label{sect:confirm}

\textbf{\texttt{Confirm}} clauses add a sequent with a single goal to the final sequent conjunction. This, and the ``assume'' rule that follows in the next section, represent an intermediate type of assertion which generally arise from the application of other (larger) proof rules. For example, applying RESOLVE's call rule generates (within the assertive code) a \texttt{\textbf{Confirm}} clause asserting that the called operation's argument-specialized precondition $\varphi$ holds. The confirm rule is given below:
\begin{align*}
\infrule{Confirm}{\mathcal{C} \backslash \ \mathit{c}; \quad \texttt{\textbf{Confirm}}  \ \bigwedge\Psi \cup \{ \vdash \varphi \}; }{\mathcal{C}  \backslash \ \mathit{c}; \quad \texttt{\textbf{Confirm}} \ \varphi; \quad \texttt{\textbf{Confirm}}\ \bigwedge\Psi;}
\end{align*}

Here and in subsequent rules we employ the turnstile $\vdash$ as a shorthand sequent constructor. For example, $\vdash \varphi$ denotes a fresh sequent with no antecedents and a single wff $\varphi$ in its succedent. In cases where we wish to add $\varphi$ to the antecedent of some non-empty sequent $s$, we will write $\Gamma_s, \varphi \vdash \Delta_s$ where $\Gamma_s$ and $\Delta_s$ refer to sets of existing wffs in the antecedent and succedent of $s$, respectively.

\begin{example}
\label{sec:runningex}
Recalling the math string notation outlined in section~\ref{sec:background}, suppose the system generated the following assertive code fragment:

\begin{lstlisting}[language=resolve]
  Assume |S| \leq 2 $\wedge$ |T| \leq 2 $\wedge$ S = \Lambda $\wedge$ T = \langle1\rangle \circ \langle2\rangle;
  Confirm (S = \Lambda $\vee$ T = \Lambda) $\wedge$ (|S| + |T| = 2);
  Confirm $\bigwedge$ $\Psi$;
\end{lstlisting}
Application of the \textsf{Confirm} rule yields:
\begin{lstlisting}[language=resolve]
  Assume |S| \leq 2 $\wedge$ |T| \leq 2 $\wedge$ S = \Lambda $\wedge$ T = \langle1\rangle \circ \langle2\rangle;
  Confirm $\bigwedge$ $\Psi \ \cup$ $\{$ {} $\vdash$ {(S = \Lambda $\vee$ T = \Lambda) $\wedge$ (|S| + |T| = 2)} $\}$;
\end{lstlisting}
The VC Generator then simplifies the final confirm with an application of \textsf{AndRight} then \textsf{OrRight} to the first (and only) sequent's succedent in the set $\Psi$, yielding:
\begin{lstlisting}[language=resolve]
  Assume |S| \leq 2 $\wedge$ |T| \leq 2 $\wedge$ S = \Lambda $\wedge$ T = \langle1\rangle \circ \langle2\rangle;
  Confirm $\bigwedge$ $\Psi \ \cup \ \{$ {} $\vdash$ {S = \Lambda, T = \Lambda} $\} \ \cup \ \{$ {} $\vdash$ {|S| + |T| = 2} $\}$;
\end{lstlisting}
In future assertive code listings, we omit the set $\Psi$ for brevity.
\end{example}

\subsection*{Parsimonious Assume Rule}
\label{sect:assumerule}

\texttt{\textbf{Assume}} clauses add antecedents to sequents in the final \texttt{\textbf{Confirm}}. The key to the new parsimonious rule, however, is to only add givens that are ``relevant'' to a given sequent. Conservative variants of our VC generator simply added each conjunct of an encountered assume clause to the list of givens for each VC---resulting in VCs with many redundant givens (e.g., Sect.~\ref{sec:motivation}). Below is our parsimonious variant of the original rule.
\begin{align*}
\infrule{ParsimoniousAssume}{\mathcal{C} \backslash \ \mathit{c}; \quad \texttt{\textbf{Confirm}} \ \bigwedge\sigma(\Psi, \varphi)}
{\mathcal{C} \backslash \ \mathit{c}; \quad \texttt{\textbf{Assume}} \ \varphi; \quad \texttt{\textbf{Confirm}} \ \bigwedge\Psi;}
\end{align*}

The core addition to our revised rule (aside from the usage of sequents) is the addition of a parsimonious ``selection'' function $\sigma$ that is applied to the set of existing sequents $\Psi$ in the final \texttt{\textbf{Confirm}} and the formula $\varphi$ that is being assumed---i.e., $\sigma(\Psi, \varphi)$. We  define this selection function as follows:
\begin{align*}
&\sigma : ((S : \wp(\mathbf{Sqnt}_\mathcal{V})) \times (\psi : \mathbf{Form}_\mathcal{V})) \rightarrow \wp(\mathbf{Sqnt}_\mathcal{V}) \ \triangleq  \\
&\qquad \{s : S \mid \forall \phi : \langle\hspace{-.085cm}\langle \psi \rangle\hspace{-.085cm}\rangle,\\
&\qquad\qquad\mathbf{if} \ \mathsf{FVC}(\phi, \langle\hspace{-.085cm}\langle \psi \rangle\hspace{-.085cm}\rangle) \cap \mathsf{SFV}_\vdash(s) \neq \emptyset \ \textbf{ then } \ s = \Gamma_s, \phi \vdash \Delta_s \ \mathbf{else} \ s = \Gamma_s \vdash \Delta_s\}.
\end{align*}

Specifically, $\sigma$ takes a set of existing sequents $S$ along with a formula $\psi$ and produces a set of (potentially modified) sequents. Here, the $\langle\hspace{-.085cm}\langle \bullet \rangle\hspace{-.085cm}\rangle$ operator is used to split $\psi$ into a list of conjuncts. The comprehension in the body of $\sigma$ tests whether or not the transitive closure of all free variables of $\phi$ appearing across of the collection of any conjuncted clauses in $\psi$ (computed via a free-var closure function $\mathsf{FVC}(\phi, \langle\hspace{-.085cm}\langle \psi \rangle\hspace{-.085cm}\rangle)$ intersects with the free variables obtained from the existing sequent $s$. If the intersection is non-empty, $\phi$ is added to the antecedent of $s$, otherwise $s$ remains unchanged. 

While the benefits of the parsimonious assume rule will not be immediately evident from example~\ref{sec:runningex}, its impact on the number of givens in VCs generated by larger programs and algorithms will be examined further in Sect.~\ref{sec:expeval}.

\begin{example}
\label{ex:example2}
To illustrate why the $\mathsf{FVC}$ function is needed, we consider the following assertive code (where each specificational free variable is subscripted by a $v$):
\begin{lstlisting}[language=resolve]
  Assume p(c$_v$) $\wedge$ c$_v$ = b$_v$ $\wedge$ s$_v$ = x$_v$
  Confirm $\bigwedge$ {} $\vdash$ {p(b$_v$)} 
\end{lstlisting}
Suppose we break apart the \texttt{Assume} clause (i.e., using $\langle\hspace{-.085cm}\langle \bullet \rangle\hspace{-.085cm}\rangle$) and consider each conjunct isolation, one at a time, from left to right. The sfv set from the first clause, \texttt{p(c}$_v$\texttt{)} (i.e.: $\{ \mathtt{c}_v\}$), does not overlap with the sfv set from the sequent in the final confirm ($\{\mathtt{b}_v\}$), thus it would not be added as an antecedent. However, the second clause \textit{would} be added to the antecedent (since \texttt{b}$_v$ overlaps). But the first clause---which has been since discarded---is needed along with the second in order to prove the goal (the third is rightfully excluded as it has no overlaps). The \textsf{FVC} closure function solves this problem by combining the sfv set of a given source clause (in this case  $\mathtt{p}(\mathtt{b}_v)$), with any other clauses in the provided list with overlapping specificational free variables. For example:
\begin{align*}
\mathsf{FVC}( \ \mathtt{p}(\mathtt{b}_v), \ \langle\hspace{-.085cm}\langle  \mathtt{p}(\mathtt{c}_v),  \ \ \mathtt{c}_v = \mathtt{b}_v, \ \ ... \ \rangle\hspace{-.085cm}\rangle \ ) = \{  \mathtt{c}_v, \mathtt{b}_v \}
\end{align*}
which enables us to add required clauses while still filtering unrelated ones.
\end{example}

\subsection*{Postprocessing Rule: Folding Top Level Equalities}

We also introduce a separate assertive code \sloppy{postprocessing/simplification} rule that performs substitutions for equalities appearing as top level formulas in a given sequent's antecedent. Here, such equalities must be of the general form $v = t$, where $v \in \mathsf{SFV}_{\mathbf{set}}(\Gamma - \{v = t\})$\footnote{The notation $\varphi[v \rightsquigarrow t]$ in the rule denotes the substitution of a variable $v$ for a term $t$ in clause $\varphi$}.
\begin{align*}
\infrule{ApplyEqLeft}{\mathcal{C} \ctxsep \ \textbf{\texttt{Confirm}} \ \bigwedge \varphi_i[v \rightsquigarrow t] \vdash \delta_j[v \rightsquigarrow t] }
{\mathcal{C} \ctxsep \ \textbf{\texttt{Confirm}} \ \bigwedge \varphi_1, \cdots, \varphi_n, \ v = t \vdash \delta_1, \cdots, \delta_m}
\textnormal{ where } \varphi_i \in \Gamma, \delta_j \in \Delta \textnormal{ and } i \leq |\Gamma|, j \leq |\Delta|
\end{align*}

Returning to example~\ref{sec:runningex}, since \textsf{ApplyEqLeft} requires $v = t$ to appear as a top-level formula in the antecedent (both of which are currently empty) and that all assertive statements prior to the final confirm clause are eliminated, the rule cannot yet be applied. Thus, we must first apply the \textsf{ParsimoniousAssume} rule which yields:

\begin{lstlisting}[language=resolve]
  Confirm $\bigwedge$ 
    {|S| \leq 2, |T| \leq 2, S = \Lambda, T = \langle1\rangle \circ \langle2\rangle} $\vdash$ {S = \Lambda, T = \Lambda}
    {|S| \leq 2, |T| \leq 2, S = \Lambda, T = \langle1\rangle \circ \langle2\rangle} $\vdash$ {|S| + |T| = 2};
\end{lstlisting}

Following this, the VC generator's simplifier then applies any applicable sequent based reduction rules,\footnote{Beyond the traditional statement rules, we broadly categorize any rule involving the $\vdash$ meta operator as a \textit{sequent} rule} which, in this case, consists of four back-to-back applications of \textsf{ApplyEqLeft} (two for each sequent), resulting in:

\begin{lstlisting}[language=resolve]
  Confirm $\bigwedge$
    {|\Lambda| \leq 2, |\langle1\rangle \circ \langle2\rangle| \leq 2} $\vdash$ {\Lambda = \Lambda, (\langle1\rangle \circ \langle2\rangle) = \Lambda}
    {|\Lambda| \leq 2, |\langle1\rangle \circ \langle2\rangle| \leq 2} $\vdash$ {|\Lambda| + |\langle1\rangle \circ \langle2\rangle| = 2};
\end{lstlisting}

The first sequent is provable via the trivially true equality: $\Lambda = \Lambda$ (recall that since each wff in the succeedent is disjuncted, this trivial equality makes the entire succeedent true). The second follows from application of various corollaries available in our theory of strings including, e.g.:
\begin{lstlisting}[language=resolve]
  Corollary Len1: \forall \alpha, \beta : SStr, |\alpha $\circ$ \beta| = |\alpha| + |\beta|;
\end{lstlisting}
as well as results involving natural numbers, \texttt{+}, and the base case of our string theory's inductively defined length $|\bullet|$ function (which establishes \texttt{|}$\Lambda$\texttt{| = 0}).

Note that while one could conceivably apply additional rules (such as reflexivity) to ``close/prove'' sequents on the fly during VC generation, we instead choose dispatch each VC only at the proving stage for consistency and reporting purposes---e.g.: displaying results after a ``prove'' button is pressed. Otherwise, a proved, missing VC may be surprising or misleading to a learner.

\subsection*{Improving Verifier Feedback for Contradictory Code}
\label{sect:stipulate}

\begin{figure}[!htb]
\centering
\includegraphics[width=\linewidth]{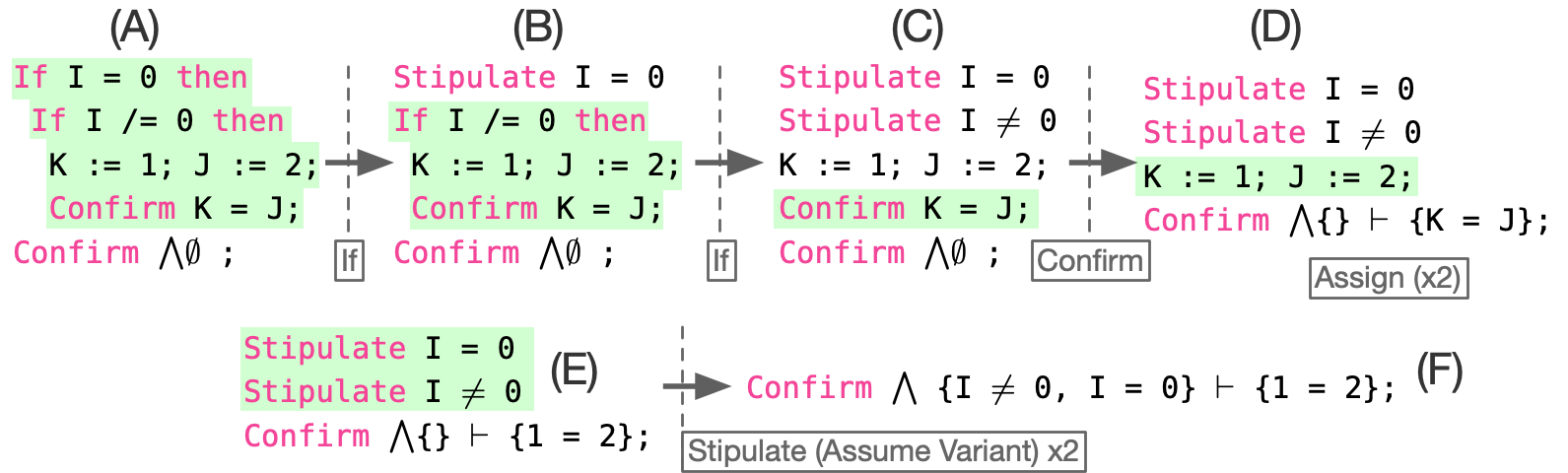}
\caption{Deriving a vacuously true sequent from contradictory nested if-statements using a variant of the \textsf{Assume} rule called \textsf{Stipulate}}
\label{fig:stipulate}
\end{figure}

The process of removing irrelevant givens requires care, as it can lead to incompleteness that negatively impacts verifier feedback. Consider, e.g, the contradictory fragment of code starting with (A) in Fig.~\ref{fig:stipulate}.

While the VC generator sets up assertive code for all nominal paths through the body, Fig.~\ref{fig:stipulate} focuses on the derivation where the innermost if-statement's body is presumed to execute. Note that each if-statement's condition is translated by the system into a \lstresolve|Stipulate| statement---meaning any control predicates resulting from the application of proof rules to \lstresolve|If|, \lstresolve|While|, etc. statements always get included as antecedents in the final confirm's sequents (regardless of variable overlap). To see why \lstresolve|Stipulate| is needed, consider step (E) in Fig.~\ref{fig:stipulate} and replace each \lstresolve|Stipulate| keyword with \lstresolve|Assume| and proceed with the derivation. Since there are no overlaps between the assumed clauses and the final confirm, each would be discarded, leaving the unprovable sequent $\vdash \{1 = 2\}$ to timeout during a proof attempt. This is not advantageous from a feedback perspective, as student code often contains similar contradictions and unreachable logic (though perhaps less obvious ones). Under this approach, vacuously true VCs can be flagged and reported to the user---identifying suspect code.

\section{Sequent-Based VC Generation in RESOLVE Studio}
\label{sec:resolvestudio}

In this section, using a sorting example, we provide an overview of the VC reporting, simplification, and derivation tracing features included in our second environment, RESOLVE Studio.

\subsection{Example: Fully Generic Sorting}

For a more complex example, we consider the specification and implementation of a generic sorting algorithm for queues. However, rather than adding a sort operation to the \lstresolve|Queue_Template| concept (which would require updating each existing implementation) we instead use an \textit{enhancement} module---which allow users to write layered code for secondary operations using the primary operations of the base concept (in this case, \lstresolve|Queue_Template|). The queue \lstresolve|Sorting_Capability| enhancement is below.
\begin{lstlisting}[language=resolve]
Enhancement Sorting_Capability (Def $\trianglelefteq$: Entry \times Entry \longrightarrow $\mathbb{B}$) for Queue_Template;
    uses Basic_Ordering_Theory,  String_Theory with Occ_Tly_Permute_Ext;
    requires Is_Total_Preordering($\trianglelefteq$);
    
    Operation Sort (updates Q : Queue);
         ensures Q Is_Permutation #Q $\wedge$ Is_Cfml_w(Q, $\trianglelefteq$);
end Sorting_Capability;
\end{lstlisting}

The enhancement is parameterized by an (abstract) binary predicate $\trianglelefteq$ that determines ordering of the queue's entries. The module level pre-condition subsequently \lstresolve|requires| that the relation passed is a total preordering---i.e.: it must be both total and transitive: 
\begin{align*}
\textnormal{($\trianglelefteq$-total)} \ \ \forall x, y : \mathtt{Entry}, \ x \trianglelefteq y \ \vee \ y \trianglelefteq x \quad \quad \textnormal{($\trianglelefteq$-trans)} \ \ \forall x, y, z : \mathtt{Entry}, \ x \trianglelefteq y \Rightarrow y \trianglelefteq z \Rightarrow x \trianglelefteq z.
\end{align*}

The \texttt{Sort} operation takes a single queue, \texttt{Q}, and \lstresolve|ensures| two properties: (1) that the resulting queue is a permutation of (exactly) the incoming queue's entries, and (2) that the ordering of all entries in the outgoing queue is conformal with (\lstresolve|Is_Cfml_w|) the $\trianglelefteq$ ordering predicate. Both the permutation and conformality predicates were imported from \lstresolve|String_Theory|, augmented with a string occurrence tallying and permutation theory extension containing the required predicates. 

For more information on the predicates that make up the sorting specification---and the way in which we define them in RESOLVE theories (including aspects of the math type system)---please consult~\cite{welch:2019}.

\subsection{An Annotated Implementation}

\begin{figure}[!htb]
\centering
\includegraphics[scale=.49]{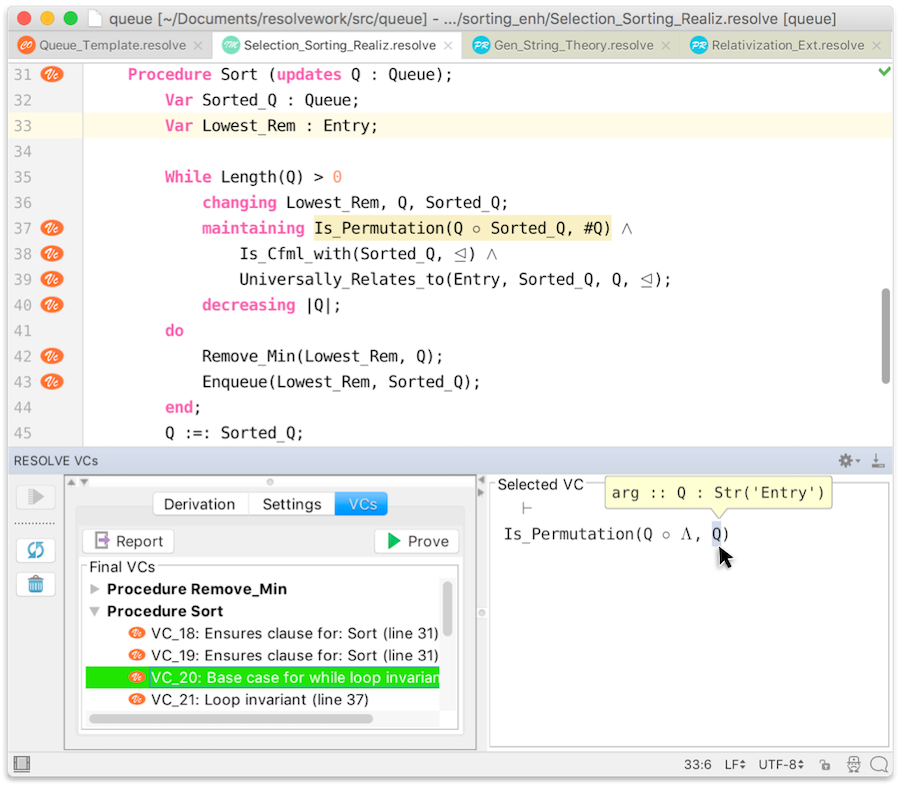}
\caption{Selection sorting impl. in RESOLVE Studio; viewing the VC for the loop invariant's base case}
\label{fig:sort}
\end{figure}

Fig.~\ref{fig:sort} shows a selection sorting implementation of \lstresolve|Sorting_Capability| loaded (with VCs already generated) in RESOLVE Studio.

Note that enhancement implementations are oblivious to any one particular implementation of the base concept, and must be written in terms of either local operations and/or by calling the primary operations of the base concept. The implementation iterates over the input queue, extracts the minimum entry in each iteration via the local procedure \lstresolve|Remove_Min|, then enqueues it onto a temporary queue, \lstresolve|Sorted_Q| (which holds the entries ordered thus far). The loop is annotated in several different parts: a \lstresolve|changing| clause indicates the variables being updated in the loop (which can help simplify certain styles of invariants), a \lstresolve|decreasing| clause (a progress metric for proving termination), and a loop invariant communicated through the \lstresolve|maintaining| clause. The invariant, shown in Fig.~\ref{fig:sort}, is summarized below:
\begin{itemize}
\item The first conjunct states that the concatenation of elements in the (temporary) \lstresolve|Sorted_Q| and \texttt{Q} constitute the entirety of the elements being sorted.
\item The second conjunct states that the \lstresolve|Sorted_Q|'s elements are ordered w.r.t. the $\trianglelefteq$ relation.
\item The last conjunct states that every element in \lstresolve|Sorted_Q| is related by $\trianglelefteq$ to every element in \texttt{Q}. In other words: that all entries in \lstresolve|Sorted_Q| ``precede'' the remaining entries in \texttt{Q}.
\end{itemize}

\subsection{Verification}
\begin{figure}[htb]
\centering
\includegraphics[width=\linewidth]{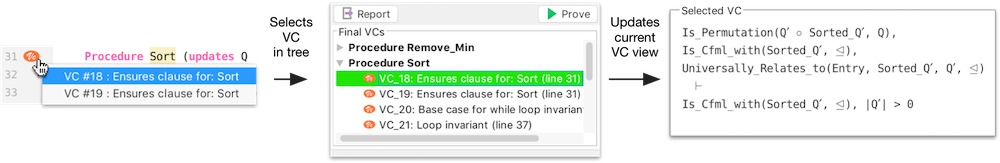}
\caption{VC selection process in RESOLVE Studio}
\label{fig:vcview}
\end{figure}

The verification process is started by invoking the ``Generate VCs'' action. 
Once VCs are generated, clicking any of the \includegraphics[scale=.45]{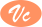} buttons opens a tool menu wherein users can select any VCs arising from that particular construct or line of code (Fig.~\ref{fig:vcview}, leftmost). 

Once a VC is selected, the IDE automatically navigates to the VC in question in the tool window, displaying the goal and givens. Selecting any particular VC in the ``Final VCs'' window also highlights the assertion, statement, or construct that generated it within the code. Pressing the prove button invokes the verifier, updating the badges with a checkmark, warning, or a timeout/failure icon.

\textbf{Advanced Feature: VC Derivation Tracing.} RESOLVE Studio also allows experienced users and researchers the ability to trace interactively through the derivation of VCs, starting from initial assertive code, to final VCs (Fig.~\ref{fig:tracing}) below.

\begin{figure}[htb]
\centering
\includegraphics[width=\linewidth]{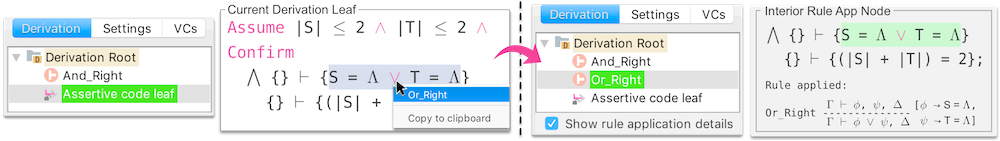}
\caption{Interactive VC derivation tracing and rule application from example~\ref{sec:runningex}}
\label{fig:tracing}
\end{figure}
We anticipate this being of use to math specialists such as theory developers who seek to add theory-specific simplification rules (e.g., simplifying $S \circ \Lambda$ to $S$) and observe their affect when applied during derivations to further `clean' resulting VCs and ease their comprehension. 

\section{Experimental Evaluation}
\label{sec:expeval}

We have tested our revised VC generation technique on a modest battery of extension operations\footnote{Recursive implementations are labeled with an $(R)$} for a variety of concepts drawn from RESOLVE's current component library including typical operations for lists, stacks, and queues.

\begin{table*}[!htb]\centering
\begin{tabular}{@{}lclllclll@{}}\toprule
\textbf{Component} & \phantom{abc}& \multicolumn{3}{c}{\textbf{Non-Parsimonious}} & \phantom{abc}& \multicolumn{3}{c}{\textbf{Parsimonious}}\\
\cmidrule{3-5} \cmidrule{7-9} 
                           && \#VCs & \#VCs$_{A \geq 5}$ & \#VCs$_{A \geq 10}$ &&  \#VCs & \#VCs$_{A \geq 5}$ & \#VCs$_{A \geq 10}$ \\ \midrule
List Search                   && 35       & 35   & 35                               && 35 & 16 & 0 \\
List Reverse (R)           && 8         & 8     & 8                                 && 8   & 0   & 0 \\
Queue Append (R)       && 8         & 8     & 8                                 && 8   & 0   & 0 \\
Queue Selection Sort  && 36        & 36   & 36                               && 36 & 16 & 0 \\
Stack Copy                  && 24        & 24   & 24                               && 24 & 5   & 0 \\
\bottomrule
\end{tabular}
\caption{Comparison of VC generation schemes: non-parsimonious vs parsimonious.}
\label{tab:vcc}
\end{table*}

Table~\ref{tab:vcc} compares VCs generated without and with the parsimonious scheme. For each, the first column contains the total number of VCs generated ($\#\mathit{VCs}$), the number of VCs generated with five or more antecedents ($\#\mathit{VCs_{A \geq 5}}$), and the number of VCs generated with ten or more antecedents ($\#\mathit{VCs_{A \geq 10}}$).

Note that with the non-parsimonious scheme, all VCs contain ten or more antecedents, while under the parsimonious scheme none contain more than ten. Since this paper has been written, the parsimonious scheme detailed has also been incorporated into the compiler backing the web-IDE, meaning students have already benefited from the reduced size of each VC. For a sense of how much shorter the VCs are, consider the (unprovable) VC from our initial example in Fig.~\ref{fig:failedvc}: $\{ 1 \leq |\langle e ''' \rangle \ \circ \ S'| \} \vdash \{S' = \langle e ''' \rangle \ \circ \ S'\}$.

These experimental results indicate that the parsimonious scheme produces fewer givens for humans to consider when debugging failed VCs. This can be viewed as a measure of \textit{usefulness} for VC comprehension overall, as well as a means of improving \textit{ease of use} when combined with our existing F-IDEs.

\section{Related Work}
\label{sec:relatedworks}

Since VC generation and auto-active program verification are vast topics, this section focuses primarily on work in generating simpler, debuggable VCs presented in an IDE or otherwise. A direct comparison of our work with other efforts is somewhat hindered by the simple fact that others typically ``outsource'' the process of generating VCs to Intermediate Verification Languages (IVLs) such as Boogie~\cite{legoues:2011}. Outsourcing has its benefits, namely: shifting the burden of VC generation (which can be non-trivial to implement in general) to a separate, reusable tool that multiple languages can target. Notable disadvantages however include high potential for ``impedance mismatches'' when translating between IVLs~\cite{ameri:2016}, or (more commonly) when translating the constructs of the rich `high level' specification language into the `lower level' representation employed by a particular IVL~\cite{utting:2017}---or vice versa~\cite{furia:2015}. These mismatches in turn can complicate error reporting efforts, including VC feedback on failed verification attempts. Some related systems are summarized below.

Dafny~\cite{leino:2013} integrates its toolchain into an IDE for Visual Studio including the Boogie Verification Debugger~\cite{legoues:2011}, which translates Z3 generated counter examples into a form suitable for human consumption. AutoProof~\cite{tschannen:2015}, which also uses Boogie and Z3 for the verification of Eiffel programs, employs ``two step" verification to broadly interpret (as opposed to on a per-VC basis) the reasons for verification failures using a combination of traditional modular verification and approximation (such as unrolling). For IDE support, push-button verification and a host of other functionality is provided in the Eiffel Verification Environment.

Why3~\cite{fillatre:2013} is another popular autoactive tool that employs its own IVL (WhyML), to target a number of SMT solvers for automated proof as well as some traditionally interactive systems (such as Coq) for proofs requiring manual steps. The language comes with a verification environment (called WhyIDE) that gives users the ability to browse goals in the current session and perform common transformations such as, for example, splitting a goal into separate conjuncts (which can then be sent to different provers). 

The KeY framework~\cite{wolfgang:2016}, which targets full functional verification of Java programs, perhaps gives users the most control over how VCs are structured and ultimately dispatched. As opposed to being reliant on any particular IVL, the language instead employs its own in-house prover that attempts to automatically verify a VC via repeated application of first order sequent reduction rules (among others). In cases where automation fails, the system can also serve as an interactive proof assistant that allows users to systematically apply `taclets' (i.e., \textit{tactics}) to the current proof state---manually guiding the system towards the goal.

\section{Conclusions and Future Work}
\label{sec:conclusion}

We have discussed the design of two F-IDEs that aim to provide users with feedback suitable for reasoning about code correctness when verification fails. Following a technical overview of our revised proof system, including rules for deriving ``parsimonious'' VCs, we have demonstrated our presentation of VCs in both the context of the web-IDE (in use for several years) and a newer F-IDE named RESOLVE Studio. A critical avenue for future work remains the ability to efficiently persist verification results to ease usability and to support scalable development of larger, more intricate case studies. Though we have extensive positive feedback from students who have used our online reasoning tools for which the VC generator detailed in this paper now forms the backbone~\cite{fowler:2021,fowler:2020,priester:2016}, we anticipate additional studies on student usage and debugging of VCs in our second experimental environment, RESOLVE Studio. 

\subsection*{Acknowledgments}
We thank the RSRG research groups at Clemson and Ohio State who have contributed to this work. Particular thanks are due to Bill Ogden and Joan Krone for their insights on the proof rules detailed.

\nocite{*}
\bibliographystyle{eptcs}
\bibliography{references}
\end{document}